\documentclass[10pt]{article}
\usepackage{graphicx}
\usepackage{amsmath}
\usepackage{amssymb}
\usepackage{caption2}
\begin{document}
\begin{center}
{\large {\bf \sc{  Analysis of the  mass and width of the $Y(4274)$ as axialvector molecule-like state
  }}} \\[2mm]
Zhi-Gang  Wang \footnote{E-mail: zgwang@aliyun.com.  }   \\
 Department of Physics, North China Electric Power University, Baoding 071003, P. R. China
\end{center}

\begin{abstract}
In this article, we assign the $Y(4274)$ to be the color octet-octet  type axialvector molecule-like state with $J^{PC}=1^{++}$ tentatively, and construct the  color octet-octet type axialvector current to study its mass and width with the QCD sum rules in details.   The predicted mass favors assigning  the   $Y(4274)$  to be  the color octet-octet  type  molecule-like  state, but the predicted width disfavors assigning  the   $Y(4274)$  to be  the color octet-octet  type  molecule-like  state strongly. The $Y(4274)$ may be the conventional  charmonium  state $\chi_{c1}(\rm 3P)$, and it is important to observe the decay $Y(4274)\to J/\psi \omega$ to diagnose the nature of the $Y(4274)$.
\end{abstract}

 PACS number: 12.39.Mk, 12.38.Lg

Key words: Molecule-like   state, QCD sum rules

\section{Introduction}

In 2011,  the CDF collaboration confirmed the
$Y(4140)$ in the $B^\pm\rightarrow J/\psi\,\phi K^\pm$ decays  with
a  statistical significance greater  than $5\,\sigma$, and observed an evidence for a second structure ($Y(4274)$) with approximate significance of $3.1\,\sigma$. The
measured mass and width
 are $\left(4274.4^{+8.4}_{-6.7}\pm1.9\right)\,\rm{MeV}$ and
$\left(32.3^{+21.9}_{-15.3}\pm7.6\right)\,\rm{MeV}$, respectively
\cite{CDF1101}. The  $Y(4274)$ may be a S-wave $D_s\bar{D}_{s0}(2317)+h.c.$ molecular state \cite{LiuLZ1011}, or not a S-wave $D_s\bar{D}_{s0}(2317)+h.c.$ molecular state \cite{Wang4274-1102}. In 2013, the CMS  collaboration  observed an evidence for a second peaking structure besides the $Y(4140)$ with the  mass
$ 4313.8 \pm 5.3 \pm 7.3 \, \rm{MeV}$ and width $ 38^{+30}_{-15}\pm  16 \,\rm{MeV}$, respectively \cite{CMS1309}.

Recently, the LHCb collaboration performed the first full amplitude analysis of the decays $B^+\to J/\psi \phi K^+$ and confirmed the two old particles $Y(4140)$ and $Y(4274)$ in the $J/\psi \phi$   mass spectrum  with statistical significances $8.4\sigma$ and $6.0\sigma$, respectively, and determined the    quantum numbers   to be $J^{PC} =1^{++}$ with statistical significances $5.7\sigma$ and $5.8\sigma$, respectively \cite{LHCb-4500-1606.07895,LHCb-4500-1606.07898}. The measured masses and widths are
\begin{flalign}
 & Y(4140) : M = 4146.5 \pm 4.5 ^{+4.6}_{-2.8} \mbox{ MeV}\, , \, \Gamma = 83 \pm 21 ^{+21}_{-14} \mbox{ MeV} \, , \nonumber\\
 & Y(4274) : M = 4273.3 \pm 8.3 ^{+17.2}_{-3.6} \mbox{ MeV}\, , \, \Gamma = 56 \pm 11 ^{+8}_{-11} \mbox{ MeV} \,  .
\end{flalign}
The LHCb collaboration determined the quantum numbers of the $Y(4274)$ to be $J^{PC}=1^{++}$, which rules  out the $0^{-+}$ molecule assignment, which is consistent with our previous work \cite{Wang4274-1102}. There have been several possible assignments, such as the color sextet-sextet type $cs\bar{c}\bar{s}$ tetraquark state \cite{Zhu-X4140,Zhu-X4274}, the conventional orbitally excited state $\chi_{c1}(\rm 3P)$ \cite{LiuXH-4274}, the color triplet-triplet type $ \frac{1}{\sqrt{6}}(u\bar{u}+d\bar{d}-2s\bar{s})c\bar{c}$ tetraquark state \cite{ZhuR-Y4274}, etc.

In 2014, the Belle collaboration analyzed the $\bar{B}^0\to K^- \pi^+ J/\psi$ decays  and observed a resonance ($Z_c(4200)$) in the $J/\psi \pi^+$ invariant mass distribution  with a statistical significance
of greater than $6.2\,\sigma$, the measured  Breit-Wigner   mass and width are
$M_{Z_c(4200)} = 4196^{+31}_{-29}{}^{+17}_{-13} \,\rm{MeV}$
and  $\Gamma_{Z_c(4200)} = 370^{+70}_{-70}{}^{+70}_{-132}\,\rm{MeV}$, respectively \cite{Zc4200exp,YuanCZ4200}.
 The preferred  spin-parity is $J^P = 1^+$.

In Ref.\cite{Wang4430-1GeV}, we study the axialvector hidden charm and hidden bottom tetraquark states in details with the QCD sum rules  and obtain the mass $M_{cu\bar{c}\bar{d},J^P=1^+}=(4.44 \pm 0.19)\,\rm{GeV}$ for the  diquark-antidiquark type tetraquark state.
In Ref.\cite{ChenZhu416}, Chen and Zhu study the vector and axialvector charmonium-like tetraquark states with the QCD sum rules in a systematic way and obtain the mass  $M_{cu\bar{c}\bar{d},J^P=1^+}=(4.16 \pm 0.10)\,\rm{GeV}$ for the diquark-antidiquark type tetraquark state. In Ref.\cite{Wang4430-1GeV}, we choose  the input parameters $m_c(\rm 1\,GeV)$, $\langle\bar{q}q\rangle(\rm 1\,GeV)$, $\langle\bar{q}g_s \sigma Gq\rangle(\rm 1\,GeV)$, while in Ref.\cite{ChenZhu416}, Chen and Zhu choose  the input parameters $m_c(m_c)$, $\langle\bar{q}q\rangle(\rm 1\,GeV)$, $\langle\bar{q}g_s \sigma Gq\rangle(\rm 1\,GeV)$.  The different predictions for the $C\gamma_5\otimes \gamma_\mu C$ type axialvector tetraquark state in Ref.\cite{Wang4430-1GeV} and Ref.\cite{ChenZhu416} originate from the different choice of the $c$-quark mass.  If we take  different choice of the heavy quark masses as a source of uncertainties, the predicted mass is about $M_{cu\bar{c}\bar{d},J^P=1^+}=(4.06-4.63)\,\rm{GeV}$.

In Ref.\cite{WangHuang3900}, we distinguish
the charge conjugations of the interpolating  currents,  study the diquark-antidiquark type axialvector  tetraquark states in a systematic way  with the QCD sum rules by taking into account the energy scale dependence of the QCD spectral densities for the first time, and obtain the predictions $M_{X(3872)}=3.87^{+0.09}_{-0.09}\,\rm{GeV}$ and $M_{Z_c(3900)}=3.91^{+0.11}_{-0.09}\,\rm{GeV}$.
In Ref.\cite{Narison1609}, R. Albuquerque et al take into account the next-to-leading order and next-to-next-to-leading order  factorizable   radiative corrections to the perturbative terms, and obtain the predication
$M_{cu\bar{c}\bar{d},J^P=1^+}=(3.888 \pm 0.130)\,\rm{GeV}$, which also depends on special choice of the energy scale $\mu$, in other words, the $\overline{MS}$ mass $m_c(\mu)$. The  non-factorizable   radiative corrections are still needed to make precise predictions. In leading order approximation, $M_{X(3872)}=3.87^{+0.09}_{-0.09}\,\rm{GeV}$ and $M_{Z_c(3900)}=3.91^{+0.11}_{-0.09}\,\rm{GeV}$ \cite{WangHuang3900}, after taking into account the next-to-leading order and next-to-next-to-leading order  factorizable   radiative corrections to the perturbative terms, $M_{cu\bar{c}\bar{d},J^P=1^+}=(3.888 \pm 0.130)\,\rm{GeV}$ \cite{Narison1609}, the predicted masses only change slightly. On the other hand, including  the next-to-leading order and next-to-next-to-leading order  factorizable   radiative corrections to the perturbative terms leads to the value of the pole residue $\lambda_{cu\bar{c}\bar{d},J^P=1^+}$ undergoes the replacement $\lambda_{cu\bar{c}\bar{d},J^P=1^+}\to 1.09\,\lambda_{cu\bar{c}\bar{d},J^P=1^+}$. According to Refs.\cite{WangHuang3900,Narison1609}, the masses of the ground state diquark-antidiquark type axialvector  tetraquark states $cu\bar{c}\bar{d}$ are about $3.9\,\rm{GeV}$.

In Ref.\cite{ChenZhu-1501}, Chen et al assign the $Z_c(4200)$ to be  the ground state axialvector tetraquark state $cu\bar{c}\bar{d}$, calculate its decay width  with the QCD sum rules, and obtain the value $\Gamma_{Z_c(4200)}=435 \pm 180\,\rm{ MeV}$. In Ref.\cite{Zhu-X4140}, Chen et al assign the $X(4140)$ to be  the ground state axialvector tetraquark state $cs\bar{c}\bar{s}$. If the $Z_c(4200)$ and $X(4140)$ are the color triplet-triplet $C\gamma_5\otimes \gamma_\mu C$ type axialvector tetraquark states, it is more natural in the case that  the $X(4140)$ has larger mass than the $Z_c(4200)$.

 In Ref.\cite{Wang-IJMPLA-4200}, we assign the $Z_c(4200)$ to be the color octet-octet type axialvector molecule-like state $\bar{u}\lambda^a c \bar{c}\lambda^a d$, where $\lambda^a$ is the Gell-Mann matrix,  and construct the  color octet-octet type axialvector current to study its mass (width) with the QCD sum rules by calculating the vacuum condensates up to dimension 10 (5) in the operator product expansion.   The predictions $M_{Z_c(4200)}=4.19 \pm 0.08\,\rm{GeV}$ and $\Gamma_{Z_c(4200)}\approx 334\,\rm{MeV}$ are consistent with the experimental data $M_{Z_c(4200)} = 4196^{+31}_{-29}{}^{+17}_{-13} \,\rm{MeV}$
and  $\Gamma_{Z_c(4200)} = 370^{+70}_{-70}{}^{+70}_{-132}\,\rm{MeV}$ from the Belle collaboration \cite{Zc4200exp,YuanCZ4200}, and favor  assigning  the $Z_c(4200)$  to be  the color octet-octet  type  molecule-like  state with $J^{PC}=1^{+-}$. Moreover, we study the energy scale dependance of the QCD spectral density of the molecule-like state in details and suggest an empirical energy scale formula to determine the ideal energy scale, in other words, to determine the  ideal $c$-quark mass.

Also in Ref.\cite{Wang-IJMPLA-4200}, we discuss the possible assignments of the $Z_c(3900)$, $Z_c(4200)$ and $Z(4430)$ as the ground state  color triplet-triplet diquark-antidiquark type tetraquark states with $J^{PC}=1^{+-}$ in details. The QCD sum rules support assigning  the $Z_c (3900)$  and $Z(4430)$  to be the ground state and the first radial excited state of the diquark-antidiquark type axialvector tetraquark states with  $J^{PC}=1^{+-}$, respectively \cite{WangHuang3900,Wang-3900-4430}.
If we assign the $Z_c(4200)$ and $Y(4274)$ to be the  molecule-like states with $J^{PC}=1^{+-}$ and $1^{++}$, respectively, the mass difference $M_{Y(4274)}-M_{Z_c(4200)}\approx 77\,\rm{MeV}$. It is reasonable,  as the $SU(3)$ breaking effects are very small for the four-quark systems \cite{Wang4430-1GeV,Wang-2016-Y4140,Wang-SU3}. In this article,  we assign the $Y(4274)$ to be the color octet-octet type molecule-like state tentatively,
\begin{eqnarray}
Y(4274)&=&\frac{1}{\sqrt{2}}\left( {\mathcal{D}}^a_s\overline{\mathcal{D}}_s^{a*} - { \mathcal{D}}_s^{a*}\overline{\mathcal{D}}^a_s\right)\,\,\,({\rm with}\,\,\,1^{++}) \, ,
\end{eqnarray}
 study its mass and decay width with the QCD sum rules in details,  where the meson-like states ${\mathcal{D}}^a_s$ and ${\mathcal{D}}_s^{a*}$ have the same quark constituents as the mesons $D_s$ and $D_s^*$ respectively, but they are in the color octet  representation, the $a$ corresponds to the Gell-Mann matrix.

The article is arranged as follows:  we derive the QCD sum rules for
the mass and width of the color octet-octet type axialvector molecule-like state  $Y(4274)$   in section 2 and in section 3 respectively;  section 4 is reserved for our conclusion.

\section{The mass of the color octet-octet type axialvector molecule-like state }
In the following, we write down  the two-point correlation function $\Pi_{\mu\nu}(p)$  in the QCD sum rules,
\begin{eqnarray}
\Pi_{\mu\nu}(p)&=&i\int d^4x e^{ip \cdot x} \langle0|T\left\{J_\mu(x)J_\nu^{\dagger}(0)\right\}|0\rangle \, , \\
J_\mu(x)&=&\frac{\bar{s}(x)i\gamma_5 \lambda^a c(x)\bar{c}(x)\gamma_\mu \lambda^a s(x)-\bar{s}(x)\gamma_\mu \lambda^a c(x)\bar{c}(x)i\gamma_5 \lambda^a s(x)}{\sqrt{2}} \, ,
\end{eqnarray}
where   the $\lambda^a$ is the Gell-Mann matrix in the color space. We construct the   color octet-octet type  current $J_\mu(x)$  to study the  molecule-like state  $Y(4274)$. One can consult  Refs.\cite{Wang-IJMPLA-4200,Wang-NPA,Wang-EPJC-8} for more literatures on the color octet-octet type  currents.
 Under charge conjugation transform $\widehat{C}$, the current $J_\mu(x)$ has the property,
\begin{eqnarray}
\widehat{C}J_{\mu}(x)\widehat{C}^{-1}&=&+ J_\mu(x) \, .
\end{eqnarray}

At the phenomenological side,  we insert  a complete set of intermediate hadronic states with
the same quantum numbers as the current operator $J_\mu(x)$ into the
correlation function $\Pi_{\mu\nu}(p)$  to obtain the hadronic representation
\cite{SVZ79,Reinders85}, and isolate the ground state
contribution,
\begin{eqnarray}
\Pi_{\mu\nu}(p)&=&\frac{\lambda_{Y(4274)}^2}{M^2_{Y(4274)}-p^2}\left(-g_{\mu\nu} +\frac{p_\mu p_\nu}{p^2}\right) +\cdots  \, ,
\end{eqnarray}
where the pole residue  $\lambda_{Y(4274)}$ is defined by $\langle 0|J_\mu(0)|Y(4274)\rangle=\lambda_{Y(4274)}\, \varepsilon_\mu$,
the $\varepsilon_\mu$ is the polarization vector of the axialvector meson $Y(4274)$.

In the following,  we briefly outline  the operator product expansion for the correlation function $\Pi_{\mu\nu}(p)$.  We contract the quark fields $s$ and $c$ in the correlation function
$\Pi_{\mu\nu}(p)$ with Wick theorem, and obtain the result,
\begin{eqnarray}
\Pi_{\mu\nu}(p)&=&-\frac{i}{2} \lambda^a_{jk}\lambda^a_{mn}\lambda^b_{k^{\prime}j^{\prime}}\lambda^b_{n^{\prime}m^{\prime}}   \int d^4x e^{ip \cdot x}   \nonumber\\
&&\left\{{\rm Tr}\left[ \gamma_5 S_c^{kk^{\prime}}(x)\gamma_5 S^{j^{\prime}j}(-x)\right] {\rm Tr}\left[ \gamma_\mu S^{nn^{\prime}}(x)\gamma_\nu S_c^{m^{\prime}m}(-x)\right] \right. \nonumber\\
&&+{\rm Tr}\left[ \gamma_\mu S_c^{kk^{\prime}}(x)\gamma_\nu S^{j^{\prime}j}(-x)\right] {\rm Tr}\left[ \gamma_5 S^{nn^{\prime}}(x)\gamma_5 S_c^{m^{\prime}m}(-x)\right] \nonumber\\
&&- {\rm Tr}\left[ \gamma_\mu S_c^{kk^{\prime}}(x)\gamma_5 S^{j^{\prime}j}(-x)\right] {\rm Tr}\left[ \gamma_5 S^{nn^{\prime}}(x)\gamma_\nu S_c^{m^{\prime}m}(-x)\right] \nonumber\\
 &&\left.- {\rm Tr}\left[ \gamma_5 S_c^{kk^{\prime}}(x)\gamma_\nu S^{j^{\prime}j}(-x)\right] {\rm Tr}\left[ \gamma_\mu S^{nn^{\prime}}(x)\gamma_5 S_c^{m^{\prime}m}(-x)\right] \right\} \, ,
\end{eqnarray}
where
\begin{eqnarray}
S^{ij}(x)&=& \frac{i\delta_{ij}\!\not\!{x}}{ 2\pi^2x^4}
-\frac{\delta_{ij}m_s}{4\pi^2x^2}-\frac{\delta_{ij}\langle
\bar{s}s\rangle}{12} +\frac{i\delta_{ij}\!\not\!{x}m_s
\langle\bar{s}s\rangle}{48}-\frac{\delta_{ij}x^2\langle \bar{s}g_s\sigma Gs\rangle}{192}+\frac{i\delta_{ij}x^2\!\not\!{x} m_s\langle \bar{s}g_s\sigma
 Gs\rangle }{1152}\nonumber\\
&& -\frac{ig_s G^{a}_{\alpha\beta}t^a_{ij}(\!\not\!{x}
\sigma^{\alpha\beta}+\sigma^{\alpha\beta} \!\not\!{x})}{32\pi^2x^2} -\frac{i\delta_{ij}x^2\!\not\!{x}g_s^2\langle \bar{s} s\rangle^2}{7776} -\frac{\delta_{ij}x^4\langle \bar{s}s \rangle\langle g_s^2 GG\rangle}{27648}-\frac{1}{8}\langle\bar{s}_j\sigma^{\mu\nu}s_i \rangle \sigma_{\mu\nu} \nonumber\\
&&   -\frac{1}{4}\langle\bar{s}_j\gamma^{\mu}s_i\rangle \gamma_{\mu }+\cdots \, ,
\end{eqnarray}
\begin{eqnarray}
S_c^{ij}(x)&=&\frac{i}{(2\pi)^4}\int d^4k e^{-ik \cdot x} \left\{
\frac{\delta_{ij}}{\!\not\!{k}-m_c}
-\frac{g_sG^n_{\alpha\beta}t^n_{ij}}{4}\frac{\sigma^{\alpha\beta}(\!\not\!{k}+m_c)+(\!\not\!{k}+m_c)
\sigma^{\alpha\beta}}{(k^2-m_c^2)^2}\right.\nonumber\\
&&\left. +\frac{g_s D_\alpha G^n_{\beta\lambda}t^n_{ij}(f^{\lambda\beta\alpha}+f^{\lambda\alpha\beta}) }{3(k^2-m_c^2)^4}
-\frac{g_s^2 (t^at^b)_{ij} G^a_{\alpha\beta}G^b_{\mu\nu}(f^{\alpha\beta\mu\nu}+f^{\alpha\mu\beta\nu}+f^{\alpha\mu\nu\beta}) }{4(k^2-m_c^2)^5}+\cdots\right\} \, , \nonumber \\
\end{eqnarray}
\begin{eqnarray}
f^{\lambda\alpha\beta}&=&(\!\not\!{k}+m_c)\gamma^\lambda(\!\not\!{k}+m_c)\gamma^\alpha(\!\not\!{k}+m_c)\gamma^\beta(\!\not\!{k}+m_c)\, ,\nonumber\\
f^{\alpha\beta\mu\nu}&=&(\!\not\!{k}+m_c)\gamma^\alpha(\!\not\!{k}+m_c)\gamma^\beta(\!\not\!{k}+m_c)\gamma^\mu(\!\not\!{k}+m_c)\gamma^\nu(\!\not\!{k}+m_c)\, ,
\end{eqnarray}
and  $t^n=\frac{\lambda^n}{2}$,  $D_\alpha=\partial_\alpha-ig_sG^n_\alpha t^n$ \cite{Reinders85}, then compute  the integrals both in the coordinate space and in the momentum space,  and obtain the correlation function $\Pi_{\mu\nu}(p)$, therefore the QCD spectral density through dispersion relation. For technical details, one can consult Ref.\cite{WangHuang3900}.

 Now  we  take the
quark-hadron duality below the continuum threshold $s_0$ and perform Borel transform  with respect to
the variable $P^2=-p^2$ to obtain  the  QCD sum rule:
\begin{eqnarray}
\lambda^2_{Y}\, \exp\left(-\frac{M^2_{Y}}{T^2}\right)= \int_{4m_c^2}^{s_0} ds\, \rho(s) \, \exp\left(-\frac{s}{T^2}\right) \, ,
\end{eqnarray}
where
\begin{eqnarray}
\rho(s)&=&\rho_{0}(s)+\rho_{3}(s) +\rho_{4}(s)+\rho_{5}(s)+\rho_{6}(s)+\rho_{7}(s) +\rho_{8}(s)+\rho_{10}(s)\, .
\end{eqnarray}
The explicit expressions of the QCD spectral densities $\rho_{0}(s)$, $\rho_{3}(s)$, $\rho_{4}(s)$, $\rho_{5}(s)$, $\rho_{6}(s)$, $\rho_{7}(s)$,
 $\rho_{8}(s)$ and $\rho_{10}(s)$ are given in the Appendix. Even in the leading order approximation, the strong coupling constant $g_s^2(\mu)=4\pi\alpha_s(\mu)$  appears according to the  equation of motion $\left(D_\alpha G_{\beta\alpha}\right)^a=g_s\sum_{q=u,d,s}\bar{q}\gamma_\beta t^aq$, see the terms $g_s^2\langle\bar{s}s\rangle^2$ in the spectral density $\rho_6(s)$. So we have to consider the energy scale dependence of the  QCD sum rules, the preferred $c$-quark mass is the $\overline{MS}$ mass $m_c(\mu)$.

 The on-shell quark propagator has no infrared divergences in perturbation theory, and this provides a perturbative definition of the quark mass  \cite{Why-pole-mass}. But the pole mass cannot be used to arbitrarily high accuracy because
of nonperturbative infrared effects in QCD. The pole mass $\widehat{m}_c$ and the $\overline{MS}$ mass $m_c(m_c)$ have the relation,
\begin{eqnarray}
\widehat{m}_c&=&m_c(m_c)\left[1+\frac{4}{3}\frac{\alpha_s(m_c)}{\pi}+\cdots \right]\, .
\end{eqnarray}
The value $m_c(m_c)=1.275 \pm 0.025\,\rm{ GeV}$ from the Particle Data Group corresponds
to $\widehat{m}_c=1.67 \pm 0.07\,\rm{ GeV}$ \cite{PDG}. If we take the pole mass, then $2\widehat{m}_c> M_{J/\psi}>M_{\eta_c}$,
at the phenomenological side of the QCD sum rules for the $J/\psi$ and $\eta_c$,
\begin{eqnarray}
&&\int_{4\widehat{m}_c^2}^{s_0} ds\, f_{J/\psi}^2 M_{J/\psi}^2\, \delta \left(s-M_{J/\psi}^2 \right) \, \exp\left(-\frac{s}{T^2} \right)=0\, , \nonumber\\
&&\int_{4\widehat{m}_c^2}^{s_0} ds\, \frac{f_{\eta_c}^2 M_{\eta_c}^2}{4\widehat{m}_c^2}\, \delta \left(s-M_{\eta_c}^2 \right) \, \exp\left(-\frac{s}{T^2} \right)=0\, .
\end{eqnarray}
If we want to obtain nonzero values, we have to choose smaller pole mass,  $2\widehat{m}_c< M_{\eta_c}<M_{J/\psi}$  \cite{WangDdecay}.
For an observable particle such as the electron, the physical mass appears as the pole mass, irrespective of the leading order, next-to-leading order, next-to-next-to-leading order, $\cdots$, radiative corrections  are concerned. In the leading order approximation, $\widehat{m}_c=m_c(m_c)$, however, the $m_c(m_c)$ originates from the  radiative corrections and  renormalization, which are beyond the leading order approximation. So in the leading order approximation, the definition of the pole mass $\widehat{m}_c$ is of arbitrary.
Again, we can see that the preferred $c$-quark mass is the $\overline{MS}$ mass $m_c(\mu)$. Moreover, the full quark
propagator has no pole because the quarks are confined. The pole mass corresponds to a non-confined particle, at the QCD side of the QCD sum rules for $J/\psi$ and $\eta_c$, the heavy quarks $c$ and $\bar{c}$ are confined particles.

 We derive   Eq.(11) with respect to  $\tau=\frac{1}{T^2}$, then eliminate the
 pole residue  $\lambda_{Y(4274)}$ to obtain the QCD sum rule for the mass,
 \begin{eqnarray}
 M_{Y}^2&=&- \frac{\int_{4m_c^2}^{s_0} ds \, \frac{d}{d \tau }\,\rho(s)e^{-\tau s}}{\int_{4m_c^2}^{s_0} ds \rho(s)e^{-\tau s}}\, .
\end{eqnarray}

Now we choose the input parameters at the QCD side of the QCD sum rules.
We take the vacuum condensates  to be the standard values
$\langle\bar{q}q \rangle=-(0.24\pm 0.01\, \rm{GeV})^3$,  $\langle\bar{s}s \rangle=(0.8\pm0.1)\langle\bar{q}q \rangle$,
 $\langle\bar{s}g_s\sigma G s \rangle=m_0^2\langle \bar{s}s \rangle$,
$m_0^2=(0.8 \pm 0.1)\,\rm{GeV}^2$, $\langle \frac{\alpha_s
GG}{\pi}\rangle=(0.33\,\rm{GeV})^4 $    at the energy scale  $\mu=1\, \rm{GeV}$
\cite{SVZ79,Reinders85,ColangeloReview}, and  take the $\overline{MS}$ masses $m_{c}(m_c)=(1.275\pm0.025)\,\rm{GeV}$ and $m_s(\mu=2\,\rm{GeV})=(0.095\pm0.005)\,\rm{GeV}$
 from the Particle Data Group \cite{PDG}.
Moreover,  we take into account
the energy-scale dependence of  the quark condensate, mixed quark condensate and $\overline{MS}$ masses from the renormalization group equation \cite{PDG,Narison-book},
 \begin{eqnarray}
 \langle\bar{s}s \rangle(\mu)&=&\langle\bar{s}s \rangle(Q)\left[\frac{\alpha_{s}(Q)}{\alpha_{s}(\mu)}\right]^{\frac{4}{9}}\, , \nonumber\\
 \langle\bar{s}g_s \sigma Gs \rangle(\mu)&=&\langle\bar{s}g_s \sigma Gs \rangle(Q)\left[\frac{\alpha_{s}(Q)}{\alpha_{s}(\mu)}\right]^{\frac{2}{27}}\, ,\nonumber\\
m_c(\mu)&=&m_c(m_c)\left[\frac{\alpha_{s}(\mu)}{\alpha_{s}(m_c)}\right]^{\frac{12}{25}} \, ,\nonumber\\
m_s(\mu)&=&m_s({\rm 2GeV} )\left[\frac{\alpha_{s}(\mu)}{\alpha_{s}({\rm 2GeV})}\right]^{\frac{4}{9}} \, ,\nonumber\\
\alpha_s(\mu)&=&\frac{1}{b_0t}\left[1-\frac{b_1}{b_0^2}\frac{\log t}{t} +\frac{b_1^2(\log^2{t}-\log{t}-1)+b_0b_2}{b_0^4t^2}\right]\, ,
\end{eqnarray}
  where $t=\log \frac{\mu^2}{\Lambda^2}$, $b_0=\frac{33-2n_f}{12\pi}$, $b_1=\frac{153-19n_f}{24\pi^2}$, $b_2=\frac{2857-\frac{5033}{9}n_f+\frac{325}{27}n_f^2}{128\pi^3}$,  $\Lambda=213\,\rm{MeV}$, $296\,\rm{MeV}$  and  $339\,\rm{MeV}$ for the flavors  $n_f=5$, $4$ and $3$, respectively  \cite{PDG}.

As the quark masses $m_c(\mu)$, $m_s(\mu)$, the quark condensate $\langle\bar{s}s \rangle(\mu)$, the mixed condensate $\langle\bar{s}g_s \sigma Gs \rangle(\mu)$ all depend on the energy scale $\mu$, the QCD spectral density  $\rho(s)$ depends on the energy scale $\mu$, we have to determine the energy scales of the QCD sum rules for those  molecule-like states in a consistent way.

The hidden charm (or bottom) four-quark systems  $q\bar{q}^{\prime}Q\bar{Q}$ can be described
by a double-well potential.     In the four-quark system $q\bar{q}^{\prime}Q\bar{Q}$,
 the heavy quark $Q$ serves as a static well potential and  combines with the light quark $q$  to form a heavy diquark $\mathcal{D}^i$  in  color antitriplet,
$q+Q \to  \mathcal{D}^i$ \cite{WangHuang3900,Wang-3900-4430,Wang-2016-Y4140,Wang-SU3, WangTetraQ},
or combines with the light antiquark $\bar{q}^\prime$ to form a heavy meson in color singlet (meson-like state in color octet),
$\bar{q}^\prime+Q \to  \bar{q}^{\prime}Q\,\, (\bar{q}^{\prime}\lambda^{a}Q)$ \cite{Wang-IJMPLA-4200,Wang-EPJC-8,WangMolecule};
 the heavy antiquark $\bar{Q}$ serves  as another static well potential and combines with the light antiquark $\bar{q}^\prime$  to form a heavy antidiquark $\overline{\mathcal{D}}^i$ in  color triplet,
$\bar{q}^{\prime}+\bar{Q} \to \overline{ \mathcal{D}}^i$ \cite{WangHuang3900,Wang-3900-4430,Wang-2016-Y4140,Wang-SU3, WangTetraQ},
or combines with the light quark $q$ to form a heavy meson in color singlet (meson-like state in color octet),
$q+\bar{Q} \to  \bar{Q}q\,\, (\bar{Q}\lambda^{a}q) $ \cite{Wang-IJMPLA-4200,Wang-EPJC-8,WangMolecule}, where the $i$ is color index, the $\lambda^a$ is Gell-Mann matrix.
 Then
\begin{eqnarray}
 \mathcal{D}^i+\overline{\mathcal{D}}^i &\to &  {\rm compact \,\,\, tetraquark \,\,\, states}\, , \nonumber\\
 \bar{q}^{\prime}Q+\bar{Q}q &\to & {\rm loose  \,\,\, molecular \,\,\, states}\, , \nonumber\\
  \bar{q}^{\prime}\lambda^aQ+\bar{Q}\lambda^a q &\to & {\rm   molecule-like  \,\,\, states}\, ,
\end{eqnarray}
the two heavy quarks $Q$ and $\bar{Q}$ stabilize the four-quark systems $q\bar{q}^{\prime}Q\bar{Q}$, just as in the case
of the $(\mu^-e^+)(\mu^+ e^-)$ molecule in QED \cite{Brodsky-2014}.

 The four-quark systems $q\bar{q}^{\prime}Q\bar{Q}$ are characterized by the effective heavy quark mass ${\mathbb{M}}_Q$ and the virtuality $V=\sqrt{M^2_{X/Y/Z}-(2{\mathbb{M}}_Q)^2}$    \cite{WangHuang3900,Wang-IJMPLA-4200,Wang-3900-4430,Wang-2016-Y4140,Wang-SU3,Wang-EPJC-8, WangTetraQ,WangMolecule}. It is natural to take the energy  scale $\mu=V$. The ${\mathbb{M}}_Q$ is just an empirical parameter to determine the optimal energy scales of the QCD spectral densities, and has no relation to the pole mass $\hat{m}_Q$ or the $\overline{MS}$ mass $m_Q(\mu)$.
 
 In Refs.\cite{WangHuang3900,Wang-IJMPLA-4200,Wang-3900-4430,Wang-2016-Y4140,Wang-SU3,Wang-EPJC-8, WangTetraQ,WangMolecule}, we observe  that there exist three universal values for the  effective heavy quark masses ${\mathbb{M}}_Q$, which correspond to  the compact   tetraquark   states, molecular   states, molecule-like    states, respectively. The empirical energy scale formula  $\mu=\sqrt{M^2_{X/Y/Z}-(2{\mathbb{M}}_Q)^2}$ works well in assigning the $X(3872)$, $Z_c(3900)$, $Y(3915)$, $Z_c(4020/4025)$, $Y(4140)$, $Z_c(4200)$, $Y(4260)$,  $Y(4360)$, $Z_c(4430)$, $X(4500)$, $Y(4630/4660)$,  $X(4700)$, $Z_b(10610)$, $Z_b(10650)$, etc.

 We evolve   all the input parameters in the QCD spectral density to the special energy scale determined by the empirical formula,
\begin{eqnarray}
\mu&=&\sqrt{M^2_{X/Y/Z}-(2{\mathbb{M}}_c)^2} \, .
 \end{eqnarray}
  In Ref.\cite{Wang-IJMPLA-4200}, we obtain the  effective mass    ${\mathbb{M}}_c=1.98\,\rm{GeV}$ for the molecule-like  states. Then we re-checked the numerical calculations  and corrected a small error concerning the mixed condensate,  the updated value is ${\mathbb{M}}_c=2.01\,\rm{GeV}$. From the empirical energy scale formula, we can obtain the  energy scale $\mu=1.45\,\rm{GeV}$. After taking into account the $SU(3)$ symmetry breaking effect $m_s-m_{u/d}\approx 0.1\,\rm{GeV}$, we obtain the optimal energy scale $\mu=1.25\,\rm{GeV}$ for the QCD spectral density $\rho(s)$. If we neglect the $SU(3)$ symmetry breaking effect, the effective
 $c$-quark mass ${\mathbb{M}}_c$  can be taken as ${\mathbb{M}}_c=2.04\,\rm{GeV}$.

Now  we search for the  Borel parameter $T^2$ and continuum threshold
parameter $s_0$  to satisfy the  following three  criteria:

$\bf{1_\cdot}$ Pole dominance at the phenomenological side;

$\bf{2_\cdot}$ Convergence of the operator product expansion;

$\bf{3_\cdot}$ Appearance of the Borel platforms.

The resulting Borel parameter and continuum threshold parameter are $T^2=(3.1-3.5)\,\rm{GeV}^2$ and $\sqrt{s_0}=(4.8\pm 0.1)\,\rm{GeV}$, respectively.
 At the Borel window, the  pole contribution is about $(41-62)\%$, the  contributions of the vacuum condensates of  dimension 8  and 10  are about $|D_8|=(5-7)\%$ and $D_{10}<1\%$, respectively, the first two criteria are satisfied.

We take into account all uncertainties of the input parameters, and obtain the values of the  mass and pole residue, which are shown explicitly in Fig.1,
 \begin{eqnarray}
 M_{Y(4274)}&=&(4.27\pm0.09) \, \rm{GeV}\, , \nonumber\\
 \lambda_{Y(4274)}&=&(4.67\pm0.74)\times 10^{-2} \, \rm{GeV}^5\, .
 \end{eqnarray}
 In Fig.1,  we plot the mass and pole residue of the $Y(4274)$    with variation of the  Borel parameter $T^2$ at a larger interval than the Borel window. From the figure, we can see that there appear platforms, the criterion $\bf{3}$  is also satisfied. Now the three  criteria are all satisfied, it is reliable to extract the ground state mass. The predicted mass $M_{Y(4274)}=(4.27\pm0.09) \, \rm{GeV}$ is consistent with the experimental value $4273.3 \pm 8.3 ^{+17.2}_{-3.6} \mbox{ MeV}$  from the LHCb collaboration \cite{LHCb-4500-1606.07895,LHCb-4500-1606.07898}, which supports assigning the $Y(4274)$ to be the color octet-octet type
 $\bar{s}\lambda^a c\bar{c}\lambda^a s$ molecule-like state.

 In Ref.\cite{Narison1609}, R. Albuquerque et al study the hidden-charm and hidden-bottom molecular states and tetraquark states  by  taking into account the next-to-leading order and next-to-next-to-leading order radiative corrections to the preturbative terms from the factorizable Feynman diagrams (without including the non-factorizable Feynman diagrams). The numerical results indicate that the predicted masses are slightly modified, while the decay constants (which relate to the pole residues) are   modified significantly, the largest modification amounts to multiplying the decay constants  by a factor $1.8$. So we expect that the  predication   $M_{Y(4274)}=(4.27\pm0.09) \, \rm{GeV}$ survives approximately  even if the next-to-leading order radiative corrections to the preturbative terms are taken into account.  Moreover, at the present time, even the  next-to-leading order factorizable contributions are not available for the color octet-octet type molecule-like states, it is a  challenging work to calculate both the
next-to-leading order factorizable and non-factorizable Feynman diagrams.

\begin{figure}
\centering
\includegraphics[totalheight=6cm,width=7cm]{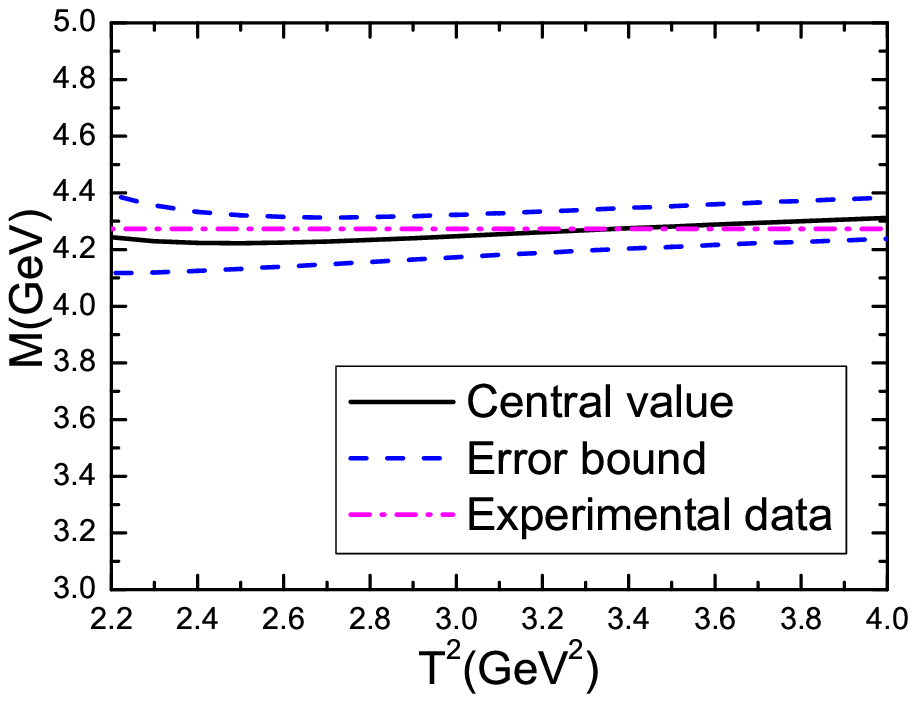}
\includegraphics[totalheight=6cm,width=7cm]{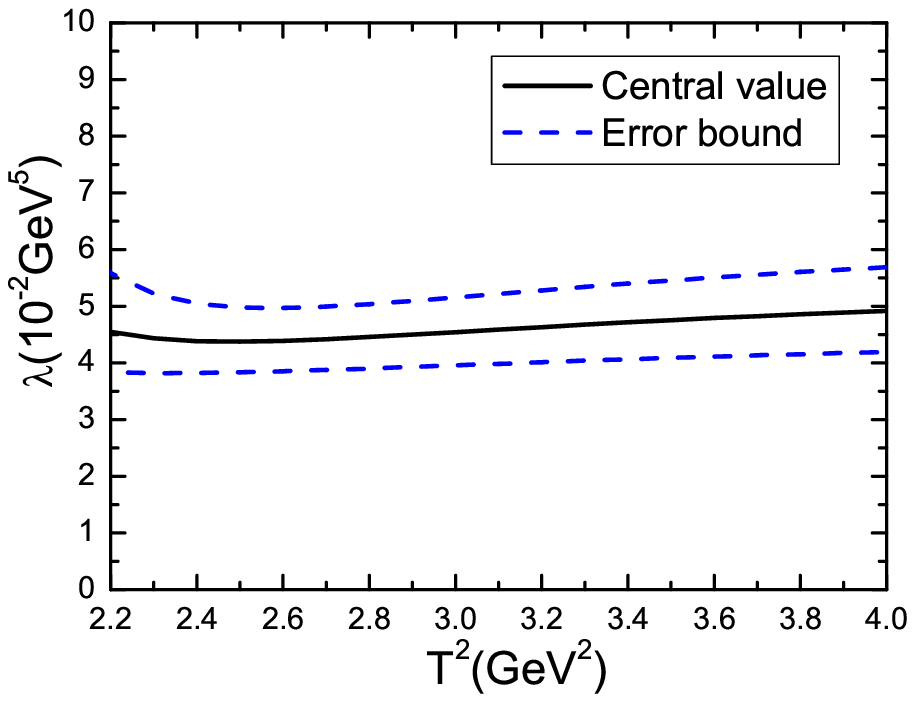}
  \caption{ The mass and pole residue of the $M_{Y(4274)}$  with variation of the  Borel parameter $T^2$. }
\end{figure}

\section{The width of the color octet-octet type axialvector molecule-like state }

We can study the strong decay $Y(4274)\to J/\psi \phi$  with the  three-point correlation function
$\Pi_{\alpha\mu\nu}(p,q)$,
\begin{eqnarray}
\Pi_{\alpha\mu\nu}(p,q)&=&i^2\int d^4xd^4y e^{ipx}e^{iqy}\langle 0|T\left\{J_\alpha^{J/\psi}(x)J_{\mu}^{\phi}(y)J_{\nu}(0)\right\}|0\rangle\, ,
\end{eqnarray}
where the currents
\begin{eqnarray}
J_\alpha^{J/\psi}(x)&=&\bar{c}(x)\gamma_\alpha c(x) \, ,\nonumber \\
J_\mu^{\phi}(x)&=&\bar{s}(y)\gamma_\mu s(y) \, ,
\end{eqnarray}
interpolate the mesons $J/\psi$ and $\phi(1020)$ according to the current-hadron couplings,
\begin{eqnarray}
\langle0|J_{\alpha}^{J/\psi}(0)|J/\psi(p)\rangle&=&f_{J/\psi}M_{J/\psi}\xi_\alpha \,\, , \nonumber \\
\langle0|J_{\mu}^{\phi}(0)|\phi(q)\rangle&=&f_{\phi}M_{\phi}\zeta_\mu \,\, ,
\end{eqnarray}
the $f_{J/\psi}$  and $f_{\phi}$ are the decay constants, the $\xi_\alpha$  and $\zeta_\mu$ are polarization vectors of the mesons $J/\psi$  and $\phi(1020)$, respectively.

At the phenomenological side,  we insert  a complete set of intermediate hadronic states with
the same quantum numbers as the current operators $J_\alpha^{J/\psi}(x)$, $J_\mu^{\phi}(y)$, $J_{\nu}(0)$ into the three-point
correlation function $\Pi_{\alpha\mu\nu}(p,q)$ and  isolate the ground state
contributions to obtain the  result,
\begin{eqnarray}
\Pi_{\alpha\mu\nu}(p,q)&=&  \frac{f_{\phi}M_{\phi} f_{J/\psi}M_{J/\psi}\lambda_{Y}G_{YJ/\psi \phi} }{(M_{Y}^2-p^{\prime2})(M_{J/\psi}^2-p^2)(M_{\phi}^2-q^2)} \,\varepsilon^{\lambda\tau\rho\theta}p^{\prime}_\lambda\left(-g_{\nu\tau}+\frac{p_{\nu}^{\prime}p^{\prime}_{\tau}}{p^{\prime 2}} \right) \left(-g_{\alpha\rho}+\frac{p_{\alpha}p_{\rho}}{p^{ 2}} \right)\nonumber\\
&&\left(-g_{\mu\theta}+\frac{q_{\mu}q_{\theta}}{q^{ 2}} \right)+\cdots  \nonumber\\
&=&\left\{ \frac{f_{\phi}M_{\phi} f_{J/\psi}M_{J/\psi}\lambda_{Y}G_{YJ/\psi \phi} }{(M_{Y}^2-p^{\prime2})(M_{J/\psi}^2-p^2)(M_{\phi}^2-q^2)} + \frac{1}{(M_{Y}^2-p^{\prime2})(M_{J/\psi}^2-p^2)} \int_{s^0_\phi}^\infty dt\frac{\rho_{Y\phi}(p^2,t,p^{\prime 2})}{t-q^2}\right.\nonumber\\
&&\left.+ \frac{1}{(M_{Y}^2-p^{\prime2})(M_{\phi}^2-q^2)} \int_{s^0_{J/\psi}}^\infty dt\frac{\rho_{YJ/\psi}(t,q^2,p^{\prime 2})}{t-p^2}+\cdots\right\}\left(\varepsilon_{\alpha\mu\nu\lambda}p^\lambda+\cdots\right) +\cdots\nonumber\\
&=&\left\{ \frac{f_{\phi}M_{\phi} f_{J/\psi}M_{J/\psi}\lambda_{Y}G_{YJ/\psi \phi} }{(M_{Y}^2-p^{\prime2})(M_{J/\psi}^2-p^2)(M_{\phi}^2-q^2)} + \frac{C_{Y\phi}}{(M_{Y}^2-p^{\prime2})(M_{J/\psi}^2-p^2)}   \right.\nonumber\\
&&\left.+ \frac{C_{YJ/\psi}}{(M_{Y}^2-p^{\prime2})(M_{\phi}^2-q^2)}  +\cdots\right\}\left(\varepsilon_{\alpha\mu\nu\lambda}p^\lambda+\cdots\right) +\cdots\, ,
\end{eqnarray}
where $p^\prime=p+q$,  the $G_{YJ/\psi\phi}$  is the hadronic coupling constant, which is defined by
\begin{eqnarray}
\langle J/\psi(p,\xi)\phi(q,\zeta)|Y(p^{\prime},\varepsilon)\rangle&=&i G_{YJ/\psi\phi} \, \varepsilon^{\lambda\tau\rho\theta} p^\prime_\lambda \varepsilon_\tau \xi_\rho \zeta_\theta\, .
\end{eqnarray}
 In this article, we choose the tensor structure $\varepsilon_{\alpha\mu\nu\lambda}p^\lambda$  to study the  coupling constant $G_{YJ/\psi\phi}$.

The two unknown functions $\rho_{Y\phi}(p^2,t,p^{\prime 2})$ and $\rho_{YJ/\psi}(t,q^2,p^{\prime 2})$ parameterize  transitions
between the ground states and the higher resonances  or the continuum states, the net effects can be  parameterized by  $C_{Y\phi}$ and $C_{YJ/\psi}$,
\begin{eqnarray}
C_{Y\phi}&=&\int_{s^0_\phi}^\infty dt\frac{ \rho_{Y\phi}(p^2,t,p^{\prime 2})}{t-q^2}\, ,\nonumber\\
C_{YJ/\psi}&=&\int_{s^0_{J/\psi}}^\infty dt\frac{\rho_{YJ/\psi}(t,q^2,p^{\prime 2})}{t-p^2}\,  .
\end{eqnarray}
In  calculations,   we take the  $C_{Y\phi}$ and $C_{YJ/\psi}$  as free parameters, and vary them  to eliminate the contaminations  to obtain  Borel platforms \cite{Ioffe-84}.

We carry out the operator product expansion up to the vacuum condensates of dimension 5 and neglect the gluon condensate, which plays a minor important role.
We obtain the QCD spectral density through dispersion relation, take the quark-hadron  duality below the continuum thresholds, then set $p^{\prime2}=p^2$ and take
double Borel transform with respect to the variables   $P^2=-p^2 $ and $Q^2=-q^2$ respectively to obtain the QCD sum rule,
\begin{eqnarray}
&& \frac{f_{\phi}M_{\phi} f_{J/\psi}M_{J/\psi}\lambda_{Y}G_{YJ/\psi \phi}}{M_{Y}^2-M_{J/\psi}^2} \left[ \exp\left(-\frac{M_{J/\psi}^2}{T_1^2} \right)-\exp\left(-\frac{M_{Y}^2}{T_1^2} \right)\right]\exp\left(-\frac{M_{\phi}^2}{T_2^2} \right) \nonumber\\
&&+C_{YJ/\psi} \exp\left(-\frac{M_{Y}^2}{T_1^2} -\frac{M_{\phi}^2}{T_2^2} \right)\nonumber\\
&&=-\frac{1}{6\sqrt{2}\pi^4}\int_{4m_c^2}^{s^0_{Y}} ds \int_{0}^{s^0_{\phi}} du  u\sqrt{1-\frac{4m_c^2}{s}}\left(m_c-\frac{m_s}{2}-\frac{m_s m_c^2}{s} \right)\exp\left(-\frac{s}{T_1^2} -\frac{u}{T_2^2} \right)\nonumber\\
&&+\frac{4m_s m_c\langle\bar{s}s\rangle}{3\sqrt{2}\pi^2} \int_{4m_c^2}^{s^0_{Y}} ds \sqrt{1-\frac{4m_c^2}{s}}\exp\left(-\frac{s}{T_1^2}  \right) \nonumber\\
&&-\frac{2\langle\bar{s}g_s\sigma Gs\rangle}{27\sqrt{2}\pi^2} \int_{4m_c^2}^{s^0_{Y}} ds \sqrt{1-\frac{4m_c^2}{s}}\frac{s+2m_c^2}{s}\exp\left(-\frac{s}{T_1^2}  \right) \nonumber\\
&&-\frac{m_s m_c\langle\bar{s}g_s\sigma Gs\rangle}{9\sqrt{2}\pi^2 T_2^2} \int_{4m_c^2}^{s^0_{Y}} ds \sqrt{1-\frac{4m_c^2}{s}}\exp\left(-\frac{s}{T_1^2}  \right) \, ,
\end{eqnarray}
where the $s^0_{Y}$ and $s^0_{\phi}$ are the continuum threshold parameters for the $Y(4274)$  and $\phi(1020)$, respectively.

The hadronic parameters are taken as    $M_{\phi}=1.019461\,\rm{GeV}$,
$M_{J/\psi}=3.0969\,\rm{GeV}$ \cite{PDG},
$f_{J/\psi}=0.418 \,\rm{GeV}$  \cite{Becirevic}, $f_{\phi}=0.253\,\rm{GeV}$, $\sqrt{s^0_{\phi}}=1.5\,\rm{GeV}$, $\sqrt{s^0_{Y}}=4.8\,\rm{GeV}$,
$M_Y=4.268\,\rm{GeV}$,   $\lambda_{Y}=4.674\times 10^{-2}\,\rm{GeV}^5$, $T_1^2=(3.1-3.5)\,\rm{GeV}^2$, $T_2^2=(2.9-3.3)\,\rm{GeV}^2$  (present work).
The unknown parameter is chosen as $C_{YJ/\psi}=0.037\,\rm{GeV}^7 $   to obtain  platforms in the Borel windows $T_1^2=(3.1-3.5)\,\rm{GeV}^2$ and $T_2^2=(2.9-3.3)\,\rm{GeV}^2$. The input parameters at the QCD side are chosen as the same in the two-point QCD sum rules for the $Y(4274)$.

\begin{figure}
\centering
\includegraphics[totalheight=6cm,width=7cm]{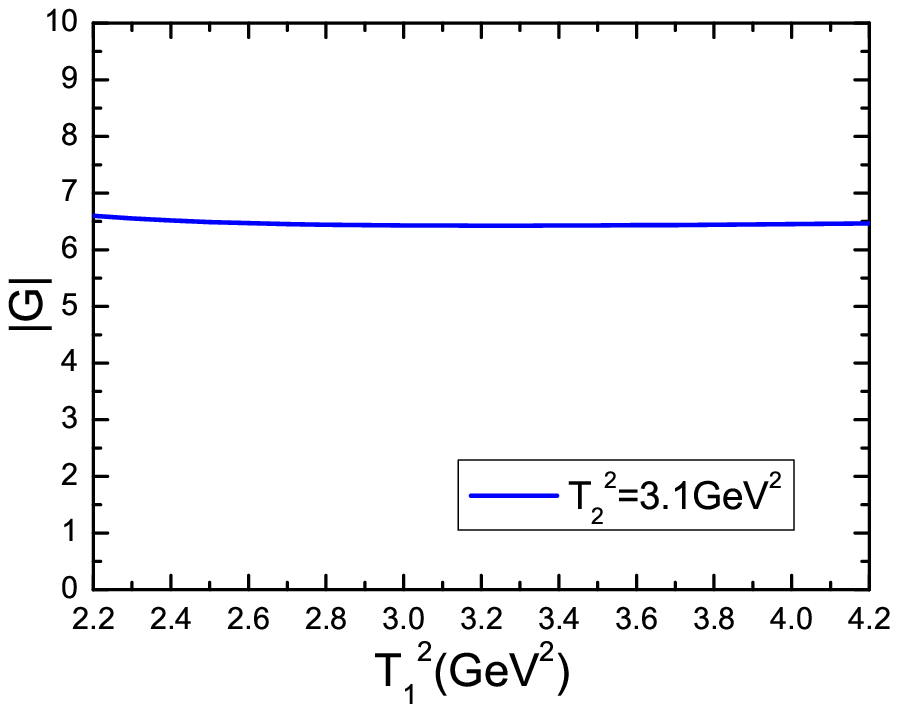}
\includegraphics[totalheight=6cm,width=7cm]{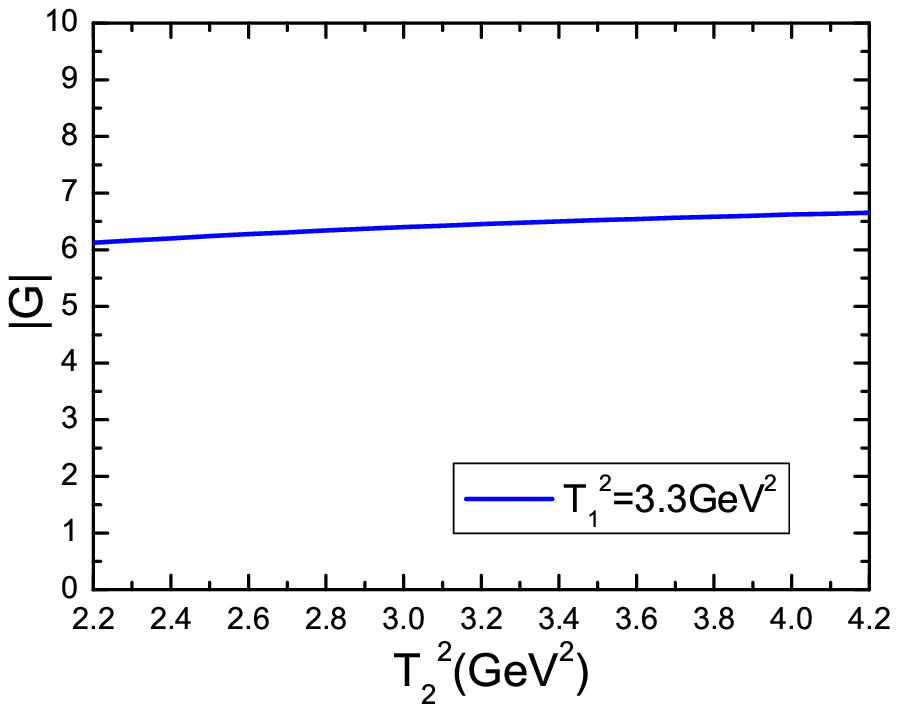}
  \caption{ The hadronic coupling constant  $G_{YJ/\psi \phi}$  with variations of the  Borel parameters $T^2_1$ and $T_2^2$, respectively. }
\end{figure}

In Fig.2, we plot the hadronic coupling constant  $G_{YJ/\psi \phi}$  with variations of the  Borel parameters $T^2_1$ and $T_2^2$, respectively. From the figure, we can see that there appear platforms in the Borel windows  $T_1^2=(3.1-3.5)\,\rm{GeV}^2$ and $T_2^2=(2.9-3.3)\,\rm{GeV}^2$, respectively. The central value of the hadronic coupling constant  $G_{YJ/\psi \phi}$ is
\begin{eqnarray}
G_{YJ/\psi \phi} &=&-6.43\, .
\end{eqnarray}
If the radiative corrections to the perturbative term of the  correlation function $\Pi_{\alpha\mu\nu}(p,q)$ also amount to multiplying a factor about $1.8$, as the color octet-octet type  current $J_\mu(x)$ is also presented, the value of the quantity  $f_{\phi} \,f_{J/\psi}\,\lambda_Y\,G_{YJ/\psi \phi}$ at the hadronic side in the QCD sum rules in Eq.(26) changes according to the rule,
\begin{eqnarray}
f_{\phi}\, f_{J/\psi}\,\lambda_Y\, G_{YJ/\psi \phi} &\to&f_{\phi}\, f_{J/\psi}\,\lambda_Y\, G_{YJ/\psi \phi}\times 1.8\, .
\end{eqnarray}
In this article, we take the values $f_{J/\psi}=0.418 \,\rm{GeV}$  \cite{Becirevic} and $f_{\phi}=0.253\,\rm{GeV}$, which include next-to-leading order radiative corrections. The factors $1.8$ come from the radiative corrections to the two-point correction function and three-point correlation function  cancel out with each other,  the net modification  of the hadronic coupling constant $G_{YJ/\psi \phi}$ is estimated to be tiny, just like the hadronic coupling constants $D^*D\pi$ and $B^*B\pi$, the net effects of the radiative corrections can be neglected \cite{DvDpion}.

Now it is easy to obtain the decay width,
\begin{eqnarray}
\Gamma(Y(4274)\to J/\psi \phi)&=& \frac{p\left(M_{Y},M_{J/\psi},M_{\phi}\right)}{24\pi M_{Y}^2}G_{YJ/\psi\phi}^2\left\{\frac{\left(M_{Y}^2-M_{\phi}^2\right)^2}{2M_{J/\psi}^2}+\frac{\left(M_{Y}^2-M_{J/\psi}^2\right)^2}{2M_{\phi}^2} \right.\nonumber\\
&&\left. +4M_Y^2-\frac{M_{J/\psi}^2+M_\phi^2}{2}\right\} \nonumber\\
&=&1.8\,{\rm{GeV}} \gg 56 \pm 11 ^{+8}_{-11} {\mbox{ MeV}}\,\,\,\rm{Experimental\,\,\,value} \,\,\cite{LHCb-4500-1606.07895,LHCb-4500-1606.07898} \, ,
\end{eqnarray}
where $p(a,b,c)=\frac{\sqrt{[a^2-(b+c)^2][a^2-(b-c)^2]}}{2a}$.	It is difficult to assign the $Y(4274)$ to be the color octet-octet type   molecule-like state $\bar{s}\lambda^a c\bar{c}\lambda^a s$. In Ref.\cite{Wang-2016-Y4140}, we assign the $Y(4140)$ to be the diquark-antidiquark type    tetraquark state $cs\bar{c}\bar{s}$ with $J^{PC}=1^{++}$,  and  study the mass and pole residue with the QCD sum rules in details. The predicted mass  disfavors assigning the $Y(4140)$ to be the $J^{PC}=1^{++}$ diquark-antidiquark type  tetraquark state $cs\bar{c}\bar{s}$. The $Y(4140)$ and $Y(4274)$ have the same quantum numbers except for the masses and widths, the QCD sum rules also disfavor assigning the $Y(4274)$ to be the $J^{PC}=1^{++}$ diquark-antidiquark type    tetraquark state $cs\bar{c}\bar{s}$.

The mass of the $\chi_{c1}(\rm 3P)$ state is $4271\,\rm{MeV}$ and $4317\,\rm{MeV}$ from the non-relativistic potential
model and the relativized Godfrey-Isgur model respectively \cite{GI2005}, which are consistent with the experimental value $4273.3 \pm 8.3 ^{+17.2}_{-3.6} \mbox{ MeV}$ from the LHCb collaboration \cite{LHCb-4500-1606.07895,LHCb-4500-1606.07898}. The width of the $\chi_{c1}(\rm 3P)$ state is $39\,\rm{MeV}$   from the non-relativistic potential model \cite{GI2005}, which is consistent with the experimental value $56 \pm 11 ^{+8}_{-11} {\mbox{ MeV}} $ from the LHCb collaboration \cite{LHCb-4500-1606.07895,LHCb-4500-1606.07898}. The $Y(4274)$ may be the conventional charmonium  $\chi_{c1}(\rm 3P)$ state \cite{LiuXH-4274}, while in Ref.\cite{ChenDY-4140}, the $Y(4140)$ is assigned to the  $\chi_{c1}(\rm 3P)$ state.  In Ref.\cite{Wang-Vector-Trans}, we study the vector meson transitions among the charmonium and bottomonium states with the heavy quark effective theory in an systematic way. If we assign the $Y(4274)$ to be the $\chi_{c1}(\rm 3P)$ state, the partial decay widths are
\begin{eqnarray}
\Gamma(Y(4274)\to J/\psi \omega)&=&17.6\times 10^{-2}{\,\rm GeV^2}\,\delta(3,1)\, , \nonumber\\
\Gamma(Y(4274)\to J/\psi \phi)&=&7.0\times 10^{-2}{\,\rm GeV^2}\, \delta(3,1)\, ,
\end{eqnarray}
where the $\delta(3,1)$ is the  hadronic coupling constant describes the transitions between the $\rm 3P$ and $\rm 1S$ charmonium multiplets \cite{Wang-Vector-Trans}.
The ratio between the two Okubo-Zweig-Iizuka suppressed decays is
\begin{eqnarray}
R&=&\frac{\Gamma(Y(4274)\to J/\psi \omega)}{\Gamma(Y(4274)\to J/\psi \phi)}=2.5\, ,
\end{eqnarray}
the decay to the final state $J/\psi \omega$ is favored due to the more available phase space. Moreover, the decay $Y(4274)\to D_s^*\bar{D}_s^*$ is Okubo-Zweig-Iizuka allowed and would have much large branching ratio. We can search for the $Y(4274)$ in the final states $J/\psi \omega$ and $D_s^*\bar{D}_s^*$ in the future.
On the other hand, if we assign the $Y(4274)$ to be the $cs\bar{c}\bar{s}$ tetraquark state, molecular state or molecule-like state, the decay $Y(4274)\to J/\psi \omega$
is doubly  Okubo-Zweig-Iizuka suppressed. It is important to observe the decay $Y(4274)\to J/\psi \omega$ to diagnose the nature of the $Y(4274)$.

\section{Conclusion}
In this article, we assign the $Y(4274)$ to be the color octet-octet  type axialvector molecule-like state with $J^{PC}=1^{++}$ tentatively, and construct the  color octet-octet type axialvector current to study its mass and width with the QCD sum rules in details.     The predicted mass $M_{Y(4274)}=(4.27\pm0.09) \, \rm{GeV}$ is consistent with the experimental value $4273.3 \pm 8.3 ^{+17.2}_{-3.6} \mbox{ MeV}$  from the LHCb collaboration, and favors  assigning the $Y(4274)$ to be the color octet-octet type  molecule-like state $\bar{s}\lambda^a c\bar{c}\lambda^a s$. The predicted width $\Gamma(Y(4274)\to J/\psi \phi)=1.8\,{\rm{GeV}}$ is much larger than the experimental value $56 \pm 11 ^{+8}_{-11} {\mbox{ MeV}}$ from the LHCb collaboration and disfavors assigning  the   $Y(4274)$  to be  the color octet-octet  type  molecule-like  state strongly. The $Y(4274)$ may be the conventional  charmonium  state $\chi_{c1}(\rm 3P)$, and the preferred decays are $Y(4274)\to D_s^*\bar{D}_s^*$, $J/\psi \omega$.  It is important to observe the decay $Y(4274)\to J/\psi \omega$ to diagnose the nature of the $Y(4274)$. We can search for the $Y(4274)$ in the final states $J/\psi \omega$ and $D_s^*\bar{D}_s^*$ in the future.

\section*{Appendix}
The explicit expressions of the QCD spectral density,
\begin{eqnarray}
\rho_{0}(s)&=&\frac{1}{1152\pi^6}\int_{y_i}^{y_f}dy \int_{z_i}^{1-y}dz \, yz\,(1-y-z)^3\left(s-\overline{m}_c^2\right)^2\left(35s^2-26s\overline{m}_c^2+3\overline{m}_c^4 \right) \nonumber\\
&&-\frac{m_s m_c}{64\pi^6}\int_{y_i}^{y_f}dy \int_{z_i}^{1-y}dz \, (y+z)\,(1-y-z)^2\left(s-\overline{m}_c^2\right)^2\left(3s-\overline{m}_c^2\right) \, ,
\end{eqnarray}

\begin{eqnarray}
\rho_{3}(s)&=&-\frac{m_c\langle \bar{s}s\rangle}{24\pi^4}\int_{y_i}^{y_f}dy \int_{z_i}^{1-y}dz \, (y+z)(1-y-z)\left(s-\overline{m}_c^2\right)\left(7s-3\overline{m}_c^2 \right) \nonumber\\
&&-\frac{m_s \langle \bar{s}s\rangle}{12\pi^4}\int_{y_i}^{y_f}dy \int_{z_i}^{1-y}dz \, yz\,(1-y-z)\left(15s^2-12s\overline{m}_c^2+3\overline{m}_c^4 \right) \nonumber\\
&&+\frac{m_s m_c^2\langle \bar{s}s\rangle}{3\pi^4}\int_{y_i}^{y_f}dy \int_{z_i}^{1-y}dz \, \left(s-\overline{m}_c^2 \right)\, ,
\end{eqnarray}

\begin{eqnarray}
\rho_{4}(s)&=&-\frac{m_c^2}{864\pi^4} \langle\frac{\alpha_s GG}{\pi}\rangle\int_{y_i}^{y_f}dy \int_{z_i}^{1-y}dz \left( \frac{z}{y^2}+\frac{y}{z^2}\right)(1-y-z)^3 \left\{ 8s-3\overline{m}_c^2+s^2\, \delta\left(s-\overline{m}_c^2\right)\right\} \nonumber\\
&&-\frac{1}{2304\pi^4}\langle\frac{\alpha_s GG}{\pi}\rangle\int_{y_i}^{y_f}dy \int_{z_i}^{1-y}dz\, (y+z)(1-y-z)^2 \,s\,(5s-4\overline{m}_c^2) \nonumber\\
&&-\frac{m_c^2}{1152\pi^4}\langle\frac{\alpha_s GG}{\pi}\rangle\int_{y_i}^{y_f}dy \int_{z_i}^{1-y}dz \left(s-\overline{m}_c^2 \right)\left\{ 7-2\left( \frac{1}{y}+ \frac{1}{z}\right) (1-y-z) \right. \nonumber\\
&&\left.+ \frac{7(1-y-z)^2}{2yz}  -\frac{7(1-y-z)}{2} +\left(\frac{1}{y}+\frac{1}{z} \right)\frac{(1-y-z)^2}{2}
 -\frac{7(1-y-z)^3}{12yz}  \right\} \nonumber\\
 && +\frac{m_s m_c^3}{192\pi^4}\langle\frac{\alpha_s GG}{\pi}\rangle\int_{y_i}^{y_f}dy \int_{z_i}^{1-y}dz\, (y+z)\left( \frac{1}{y^3}+\frac{1}{z^3}\right)(1-y-z)^2
 \left\{ 1+\frac{2s}{3}\, \delta\left(s-\overline{m}_c^2\right)\right\} \nonumber\\
&& -\frac{m_s m_c }{1152\pi^4} \langle\frac{\alpha_s GG}{\pi}\rangle\int_{y_i}^{y_f}dy \int_{z_i}^{1-y}dz \, (1-y-z) \left[1-\frac{7}{4}\left( \frac{1}{y}+\frac{1}{z}\right)(1-y-z) \right.\nonumber\\
&&\left.+ 6\left( \frac{z}{y^2}+\frac{y}{z^2}\right)(1-y-z)\right] \left( 5s-3\overline{m}_c^2\right) \, ,
\end{eqnarray}

\begin{eqnarray}
\rho_{5}(s)&=&\frac{m_c\langle \bar{s}g_s\sigma Gs\rangle}{48\pi^4}\int_{y_i}^{y_f}dy \int_{z_i}^{1-y}dz \,  (y+z) \left(5s-3\overline{m}_c^2 \right) \nonumber\\
&&+\frac{m_c\langle \bar{s}g_s\sigma Gs\rangle}{192\pi^4}\int_{y_i}^{y_f}dy \int_{z_i}^{1-y}dz   \left(\frac{y}{z}+\frac{z}{y} \right)(1-y-z) \left(2s-\overline{m}_c^2 \right)  \nonumber\\
&&-\frac{m_c\langle \bar{s}g_s\sigma Gs\rangle}{576\pi^4}\int_{y_i}^{y_f}dy \int_{z_i}^{1-y}dz   \left(\frac{y}{z}+\frac{z}{y} \right)(1-y-z) \left(5s-3\overline{m}_c^2 \right) \nonumber\\
&&+\frac{m_s\langle \bar{s}g_s\sigma Gs\rangle}{36\pi^4}\int_{y_i}^{y_f}dy \int_{z_i}^{1-y}dz   \,yz \left\{ 8s-3\overline{m}_c^2 +s^2 \delta\left(s-\overline{m}_c^2 \right)\right\} \nonumber\\
&&-\frac{m_s m_c^2\langle \bar{s}g_s\sigma Gs\rangle}{12\pi^4}\int_{y_i}^{y_f}dy  \nonumber\\
&&-\frac{m_s m_c^2\langle \bar{s}g_s\sigma Gs\rangle}{192\pi^4}\int_{y_i}^{y_f}dy \int_{z_i}^{1-y}dz   \left(\frac{1}{y}+\frac{1}{z} \right) \nonumber\\
&&+\frac{m_s\langle \bar{s}g_s\sigma Gs\rangle}{576\pi^4}\int_{y_i}^{y_f}dy \int_{z_i}^{1-y}dz\,   \left(y+z \right) \left(5s-3\overline{m}_c^2 \right) \, ,
\end{eqnarray}

\begin{eqnarray}
\rho_{6}(s)&=&\frac{2m_c^2\langle\bar{s}s\rangle^2}{9\pi^2}\int_{y_i}^{y_f}dy +\frac{g_s^2\langle\bar{s}s\rangle^2}{243\pi^4}\int_{y_i}^{y_f}dy \int_{z_i}^{1-y}dz\, yz \left\{8s-3\overline{m}_c^2 +s^2\, \delta\left(s-\overline{m}_c^2 \right)\right\}\nonumber\\
&&+\frac{g_s^2\langle\bar{s}s\rangle^2}{1296\pi^4}\int_{y_i}^{y_f}dy \int_{z_i}^{1-y}dz\, (1-y-z)\left\{ \left(\frac{z}{y}+\frac{y}{z} \right)\left(7s-4\overline{m}_c^2 \right)\right.\nonumber\\
&&\left.+\frac{1}{3}\left(\frac{z}{y^2}+\frac{y}{z^2} \right)m_c^2\left[ 7+5s\,\delta\left(s-\overline{m}_c^2 \right)\right]-\frac{1}{3}(y+z)\left(4s-3\overline{m}_c^2 \right)\right\} \nonumber\\
&&-\frac{g_s^2\langle\bar{s}s\rangle^2}{1944\pi^4}\int_{y_i}^{y_f}dy \int_{z_i}^{1-y}dz\,(1-y-z)\left\{  3\left(\frac{z}{y}+\frac{y}{z} \right)\left(2s-\overline{m}_c^2 \right)\right. \nonumber\\
&&\left.+\left(\frac{z}{y^2}+\frac{y}{z^2} \right)m_c^2\left[ 1+s\,\delta\left(s-\overline{m}_c^2\right)\right]+2(y+z)\left[8s-3\overline{m}_c^2 +s^2\, \delta\left(s-\overline{m}_c^2\right)\right]\right\}\nonumber\\
&&+\frac{m_s m_c\langle\bar{s}s\rangle^2}{6\pi^2}\int_{y_i}^{y_f}dy \left\{ 1+\frac{2s}{3}\delta\left(s-\widetilde{m}_c^2 \right)\right\} \, ,
\end{eqnarray}

\begin{eqnarray}
\rho_7(s)&=&\frac{m_c^3\langle\bar{s}s\rangle}{432\pi^2}\langle\frac{\alpha_sGG}{\pi}\rangle\int_{y_i}^{y_f}dy \int_{z_i}^{1-y}dz \,\left(y+z \right) \left(\frac{1}{y^3}+\frac{1}{z^3}\right)(1-y-z) \left( 1+\frac{2s}{T^2}\right)\delta\left(s-\overline{m}_c^2\right)\nonumber\\
&&-\frac{m_c\langle\bar{s}s\rangle}{288\pi^2}\langle\frac{\alpha_sGG}{\pi}\rangle\int_{y_i}^{y_f}dy \int_{z_i}^{1-y}dz\left\{1-7\left(\frac{1}{y}+\frac{1}{z}\right)\frac{1-y-z}{2}+12(1-y-z)\right.\nonumber\\
&&\left.\left(\frac{y}{z^2}+\frac{z}{y^2}\right)\right\} \left\{1+\frac{2s}{3}\delta\left(s-\overline{m}_c^2\right) \right\} \nonumber\\
&&-\frac{m_c\langle\bar{s}s\rangle}{144\pi^2}\langle\frac{\alpha_sGG}{\pi}\rangle\int_{y_i}^{y_f}dy \left\{1+\frac{2s}{3}\delta\left(s-\widetilde{m}_c^2\right) \right\} \nonumber\\
&&+\frac{m_s m_c^2\langle\bar{s}s\rangle}{108\pi^2T^2}\langle\frac{\alpha_sGG}{\pi}\rangle\int_{y_i}^{y_f}dy \int_{z_i}^{1-y}dz \left(\frac{y}{z^2}+\frac{z}{y^2}\right)(1-y-z)\left(s+\frac{s^2}{T^2}\right)\delta\left(s-\overline{m}_c^2\right)  \nonumber\\
&&-\frac{m_s m_c^4\langle\bar{s}s\rangle}{54\pi^2T^2}\langle\frac{\alpha_sGG}{\pi}\rangle\int_{y_i}^{y_f}dy \int_{z_i}^{1-y}dz \left(\frac{1}{y^3}+\frac{1}{z^3}\right)\delta\left(s-\overline{m}_c^2\right)  \nonumber\\
&&+\frac{m_s m_c^2\langle\bar{s}s\rangle}{18\pi^2}\langle\frac{\alpha_sGG}{\pi}\rangle\int_{y_i}^{y_f}dy \int_{z_i}^{1-y}dz \left(\frac{1}{y^2}+\frac{1}{z^2}\right)\delta\left(s-\overline{m}_c^2\right)  \nonumber\\
&&+\frac{m_s \langle\bar{s}s\rangle}{864\pi^2}\langle\frac{\alpha_sGG}{\pi}\rangle\int_{y_i}^{y_f}dy \int_{z_i}^{1-y}dz \,(y+z)\left(s+\frac{s^2}{2T^2}\right)\delta\left(s-\overline{m}_c^2\right)  \nonumber\\
&&-\frac{m_s m_c^2\langle\bar{s}s\rangle}{864\pi^2}\langle\frac{\alpha_sGG}{\pi}\rangle\int_{0}^{1}dy  \left(\frac{1}{y}+\frac{1}{1-y}\right)\delta\left(s-\widetilde{m}_c^2\right)  \nonumber\\
&&+\frac{m_s m_c^2 \langle\bar{s}s\rangle}{1728\pi^2}\langle\frac{\alpha_sGG}{\pi}\rangle\int_{y_i}^{y_f}dy \int_{z_i}^{1-y}dz \left\{ \frac{1}{y^2}+\frac{1}{z^2}+\frac{14}{yz}-\frac{7(1-y-z)}{yz}\right\}\delta\left(s-\overline{m}_c^2\right)  \nonumber\\
&&-\frac{7m_s  \langle\bar{s}s\rangle}{288\pi^2}\langle\frac{\alpha_sGG}{\pi}\rangle\int_{y_i}^{y_f}dy \int_{z_i}^{1-y}dz \left\{ 1+\frac{2s}{3}\delta\left(s-\overline{m}_c^2\right) \right\}  \nonumber\\
&&+\frac{m_s m_c^2  \langle\bar{s}s\rangle}{108\pi^2}\langle\frac{\alpha_sGG}{\pi}\rangle\int_{0}^{1}dy   \left( 1+\frac{s}{T^2}\right)\delta\left(s-\widetilde{m}_c^2\right)  \, ,
\end{eqnarray}

\begin{eqnarray}
\rho_8(s)&=&-\frac{m_c^2\langle\bar{s}s\rangle\langle\bar{s}g_s\sigma Gs\rangle}{9\pi^2}\int_0^1 dy \left(1+\frac{s}{T^2} \right)\delta\left(s-\widetilde{m}_c^2\right)\nonumber\\
&&-\frac{m_c^2\langle\bar{s}s\rangle\langle\bar{s}g_s\sigma Gs\rangle}{144\pi^2}\int_0^1 dy \left( \frac{1}{y}+\frac{1}{1-y} \right)\delta\left(s-\widetilde{m}_c^2\right)\nonumber\\
&&+\frac{\langle\bar{s}s\rangle\langle\bar{s}g_s\sigma Gs\rangle}{144\pi^2}\int_{y_i}^{y_f} dy \left\{1+\frac{2s}{3}\delta\left(s-\widetilde{m}_c^2\right) \right\}  \nonumber\\
&&-\frac{5 m_s m_c\langle\bar{s}s\rangle\langle\bar{s}g_s\sigma Gs\rangle}{108\pi^2}\int_0^1 dy \left( 1+\frac{3s}{2T^2}+\frac{s^2}{T^4} \right)\delta\left(s-\widetilde{m}_c^2\right)\nonumber\\
&&-\frac{ m_s m_c\langle\bar{s}s\rangle\langle\bar{s}g_s\sigma Gs\rangle}{288\pi^2 T^2}\int_0^1 dy \left( \frac{1-y}{y}+\frac{y}{1-y} \right)\,s\,\delta\left(s-\widetilde{m}_c^2\right)\nonumber\\
&&+\frac{ m_s m_c\langle\bar{s}s\rangle\langle\bar{s}g_s\sigma Gs\rangle}{864\pi^2  }\int_0^1 dy \left( \frac{1-y}{y}+\frac{y}{1-y} \right)\left( 1+\frac{2s}{T^2} \right)\delta\left(s-\widetilde{m}_c^2\right)\, ,
\end{eqnarray}

\begin{eqnarray}
\rho_{10}(s)&=&\frac{m_c^2\langle\bar{s}g_s\sigma Gs\rangle^2}{72\pi^2T^6}\int_0^1 dy \,s^2\,\delta \left( s-\widetilde{m}_c^2\right)\nonumber\\
&&-\frac{m_c^4\langle\bar{s}s\rangle^2}{81T^4}\langle\frac{\alpha_sGG}{\pi}\rangle\int_0^1 dy  \left\{ \frac{1}{y^3}+\frac{1}{(1-y)^3}\right\} \delta\left( s-\widetilde{m}_c^2\right)\nonumber\\
&&+\frac{m_c^2\langle\bar{s}s\rangle^2}{27T^2}\langle\frac{\alpha_sGG}{\pi}\rangle\int_0^1 dy  \left\{ \frac{1}{y^2}+\frac{1}{(1-y)^2}\right\} \delta\left( s-\widetilde{m}_c^2\right)\nonumber\\
&&-7\frac{\langle\bar{s}s\rangle^2}{1296}\langle\frac{\alpha_sGG}{\pi}\rangle\int_0^1 dy  \left( 1+\frac{2s}{T^2}\right) \delta\left( s-\widetilde{m}_c^2\right)\nonumber\\
&&+\frac{m_c^2\langle\bar{s}g_s\sigma Gs\rangle^2}{576\pi^2T^4}\int_0^1 dy \left( \frac{1}{y}+\frac{1}{1-y}\right)\,s \, \delta \left( s-\widetilde{m}_c^2\right)\nonumber\\
&&-\frac{\langle\bar{s}g_s\sigma Gs\rangle^2}{864\pi^2}\int_0^1 dy \left(1+\frac{3s}{2T^2}+\frac{s^2}{T^4} \right)\delta \left( s-\widetilde{m}_c^2\right)\nonumber\\
&&-\frac{\langle\bar{s}g_s\sigma Gs\rangle^2}{5832\pi^2}\int_0^1 dy \left(1+\frac{2s}{T^2}  \right)\delta \left( s-\widetilde{m}_c^2\right)\nonumber\\
&&+\frac{m_c^2\langle\bar{s}s\rangle^2}{81T^6}\langle\frac{\alpha_sGG}{\pi}\rangle\int_0^1 dy \, s^2 \, \delta\left( s-\widetilde{m}_c^2\right) \nonumber\\
&&-\frac{ m_s m_c\langle\bar{s}g_s\sigma Gs\rangle^2}{216\pi^2T^2}\int_0^1 dy \left(1+\frac{s}{T^2}+\frac{s^2}{2T^4} -\frac{s^3}{T^6}\right)\delta \left( s-\widetilde{m}_c^2\right)\nonumber\\
&&+\frac{m_s m_c^3\langle\bar{s}s\rangle^2}{108T^4}\langle\frac{\alpha_sGG}{\pi}\rangle\int_0^1 dy  \left\{ \frac{1}{y^3}+\frac{1}{(1-y)^3}\right\}\left( 1-\frac{2s}{3T^2}\right) \delta\left( s-\widetilde{m}_c^2\right)\nonumber\\
&&-\frac{m_s m_c \langle\bar{s}s\rangle^2}{108T^2}\langle\frac{\alpha_sGG}{\pi}\rangle\int_0^1 dy  \left\{ \frac{1-y}{y^2}+\frac{y}{(1-y)^2}\right\}\left( 1-\frac{2s}{T^2}\right) \delta\left( s-\widetilde{m}_c^2\right)\nonumber\\
&&+\frac{7m_s m_c \langle\bar{s}s\rangle^2}{2592T^2}\langle\frac{\alpha_sGG}{\pi}\rangle\int_0^1 dy  \left( \frac{1}{y}+\frac{1}{1-y}\right)\left( 1-\frac{2s}{T^2}\right) \delta\left( s-\widetilde{m}_c^2\right)\nonumber\\
&&-\frac{ m_s m_c\langle\bar{s}g_s\sigma Gs\rangle^2}{1728\pi^2T^2}\int_0^1 dy \, \left( \frac{1-y}{y}+\frac{y}{1-y}\right)\left(1+\frac{s}{T^2}-\frac{s^2}{T^4} \right)\delta \left( s-\widetilde{m}_c^2\right)\nonumber\\
&&+\frac{ m_s m_c\langle\bar{s}g_s\sigma Gs\rangle^2}{5184\pi^2T^2}\int_0^1 dy \, \left( \frac{1-y}{y}+\frac{y}{1-y}\right)\left(1+\frac{s}{T^2}-\frac{2s^2}{T^4} \right)\delta \left( s-\widetilde{m}_c^2\right)\nonumber\\
&&-\frac{m_s m_c \langle\bar{s}s\rangle^2}{324T^2}\langle\frac{\alpha_sGG}{\pi}\rangle\int_0^1 dy \,\left( 1+\frac{s}{T^2}+\frac{s^2}{2T^4}-\frac{s^3}{T^6}\right) \delta\left( s-\widetilde{m}_c^2\right)\, ,
\end{eqnarray}
the subscripts  $0$, $3$, $4$, $5$, $6$, $7$, $8$, $10$ denote the dimensions of the  vacuum condensates; $y_{f}=\frac{1+\sqrt{1-4m_c^2/s}}{2}$,
$y_{i}=\frac{1-\sqrt{1-4m_c^2/s}}{2}$, $z_{i}=\frac{ym_c^2}{y s -m_c^2}$, $\overline{m}_c^2=\frac{(y+z)m_c^2}{yz}$,
$ \widetilde{m}_c^2=\frac{m_c^2}{y(1-y)}$, $\int_{y_i}^{y_f}dy \to \int_{0}^{1}dy$, $\int_{z_i}^{1-y}dz \to \int_{0}^{1-y}dz$ when the $\delta$ functions $\delta\left(s-\overline{m}_c^2\right)$ and $\delta\left(s-\widetilde{m}_c^2\right)$ appear.

\section*{Acknowledgements}
This  work is supported by National Natural Science Foundation,
Grant Number 11375063, and Natural Science Foundation of Hebei province, Grant Number A2014502017.


\begin{thebibliography}{99}

\bibitem{CDF1101} T. Aaltonen et al,  arXiv:1101.6058.

\bibitem{LiuLZ1011} X. Liu, Z. G. Luo and S. L. Zhu, Phys. Lett. {\bf B699} (2011) 341;
 J. He and X. Liu,  Eur. Phys. J. {\bf C72} (2012) 1986;
 S. I. Finazzo, X. Liu and M. Nielsen, Phys. Lett. {\bf B701} (2011) 101.

\bibitem{Wang4274-1102}  Z. G. Wang, Int. J. Mod. Phys. {\bf A26} (2011) 4929.

\bibitem{CMS1309} S. Chatrchyan  et al,  Phys. Lett. {\bf B734} (2014) 261.

\bibitem{LHCb-4500-1606.07895}  R. Aaij et al,  Phys. Rev. Lett. {\bf 118} (2017) 022003.

\bibitem{LHCb-4500-1606.07898}  R. Aaij et al,  Phys. Rev. {\bf D95} (2017)  012002.


\bibitem{Zhu-X4140} H. X. Chen, E. L. Cui, W. Chen, X. Liu and S. L. Zhu, arXiv:1606.03179.

\bibitem{Zhu-X4274} J. Wu, Y. R. Liu, K. Chen, X. Liu and S. L. Zhu, Phys. Rev. {\bf D94} (2016) 094031.

\bibitem{LiuXH-4274} X. H. Liu, arXiv:1607.01385;
Q. F. Lu and Y. B. Dong, Phys. Rev. {\bf D94} (2016)  074007.

\bibitem{ZhuR-Y4274} R. Zhu,  Phys. Rev. {\bf D94} (2016)  054009.



\bibitem{Zc4200exp} K. Chilikin et al, Phys. Rev. {\bf D90} (2014) 112009.

\bibitem{YuanCZ4200} C. Z. Yuan, Int. J. Mod. Phys. {\bf A29} (2014) 1430046.

\bibitem{Wang4430-1GeV}  Z. G. Wang, Eur. Phys. J. {\bf C70} (2010) 139.

\bibitem{ChenZhu416} W. Chen and S. L. Zhu, Phys. Rev. {\bf D83} (2011) 034010.

\bibitem{WangHuang3900}  Z. G. Wang and T. Huang,  Phys. Rev. {\bf D89} (2014)  054019.


\bibitem{Narison1609} R. Albuquerque, S. Narison, F. Fanomezana, A. Rabemananjara, D. Rabetiarivony and G. Randriamanatrika,
Int. J. Mod. Phys. {\bf A31} (2016) 1650196.


\bibitem{ChenZhu-1501} W. Chen, T. G. Steele, H. X. Chen and S. L. Zhu, Eur. Phys. J. {\bf C75} (2015)  358.




\bibitem{Wang-IJMPLA-4200} Z. G. Wang, Int. J. Mod. Phys. {\bf A30} (2015)  1550168.

\bibitem{Wang-3900-4430} Z. G. Wang, Commun. Theor. Phys. {\bf 63} (2015) 325.




\bibitem{Wang-2016-Y4140} Z. G. Wang, Eur. Phys. J. {\bf C76} (2016) 657.

\bibitem{Wang-SU3}  Z. G. Wang, Eur. Phys. J. {\bf C74} (2014) 2874.


\bibitem{Wang-NPA}  Z. G. Wang, Nucl. Phys. {\bf A791} (2007) 106;
L. Tang and C. F. Qiao, Eur. Phys. J. {\bf C76} (2016)  558.


\bibitem{Wang-EPJC-8} Z. G. Wang and T. Huang, Eur. Phys. J. {\bf C74} (2014)  2891.

\bibitem{SVZ79} M. A. Shifman, A. I. Vainshtein and V. I. Zakharov, Nucl. Phys. {\bf B147} (1979) 385; Nucl. Phys. {\bf B147} (1979) 448.

\bibitem{Reinders85} L. J. Reinders, H. Rubinstein and S. Yazaki, Phys. Rept. {\bf 127} (1985) 1.


\bibitem{Why-pole-mass} R. Tarrach, Nucl. Phys. {\bf B183} (1981) 384;
A. Kronfeld, Phys. Rev. {\bf D58} (1998) 051501.

\bibitem{PDG}   K. A. Olive et al, Chin. Phys. {\bf C38} (2014) 090001.

\bibitem{WangDdecay} Z. G. Wang, JHEP {\bf 1310} (2013) 208;
Z. G. Wang, Eur. Phys. J. {\bf C75} (2015) 427.


\bibitem{ColangeloReview} P. Colangelo and A. Khodjamirian, hep-ph/0010175.


\bibitem{Narison-book} S. Narison, Camb. Monogr. Part. Phys. Nucl. Phys. Cosmol. {\bf 17} (2002) 1.

\bibitem{WangTetraQ} Z. G. Wang and T. Huang, Nucl. Phys. {\bf A930} (2014) 63;
Z. G. Wang, Commun. Theor. Phys. {\bf 63} (2015)  466;
Z. G. Wang and Y. F. Tian, Int. J. Mod. Phys. {\bf A30} (2015) 1550004;
Z. G. Wang, Eur. Phys. J. {\bf C76} (2016)  387;
Z. G. Wang, Eur. Phys. J. {\bf C77} (2017)  78.

\bibitem{WangMolecule} Z. G. Wang, Eur. Phys. J. {\bf C74} (2014) 2963.




\bibitem{Brodsky-2014}  S. J. Brodsky, D. S. Hwang and R. F. Lebed, Phys. Rev. Lett. {\bf 113} (2014) 112001.


\bibitem{Ioffe-84} B. L. Ioffe and A. V. Smilga, Nucl. Phys. {\bf B232} (1984) 109;
 Z. G. Wang, W. M. Yang and S. L. Wan, Phys. Rev. {\bf D72} (2005) 034012.


\bibitem{Becirevic} D. Becirevic, G. Duplancic, B. Klajn, B. Melic and F. Sanfilippo,  Nucl. Phys. {\bf B883} (2014) 306.

\bibitem{DvDpion} A. Khodjamirian, R. Ruckl, S. Weinzierl and O. I. Yakovlev, Phys. Lett. {\bf B457} (1999) 245.


\bibitem{GI2005} T. Barnes, S. Godfrey and E. S. Swanson, Phys. Rev. {\bf D72} (2005) 054026.

\bibitem{ChenDY-4140} D. Y. Chen,   Eur. Phys. J. {\bf C76} (2016) 671.


\bibitem{Wang-Vector-Trans} Z. G. Wang, Commun. Theor. Phys. {\bf 57} (2012) 93.

\end{thebibliography}
\end{document}